\documentclass{JHEP3}
\usepackage{amssymb}
\usepackage{graphicx}
\usepackage{epsfig}
\newcommand{\bmat}{\left(\begin{array}}
\newcommand{\emat}{\end{array}\right)}
\newcommand{\be}{\begin{equation}}
\newcommand{\ee}{\end{equation}}
\newcommand{\bea}{\begin{eqnarray}}
\newcommand{\eea}{\end{eqnarray}}
\newcommand{\nn}{\nonumber}

\def\L{\left}
\def\R{\right}

\def\lsim{\raise0.3ex\hbox{$\;<$\kern-0.75em\raise-1.1ex\hbox{$\sim\;$}}}
\def\gsim{\raise0.3ex\hbox{$\;>$\kern-0.75em\raise-1.1ex\hbox{$\sim\;$}}}

\date{\today}
\title{Dark Matter Annihilation and the PAMELA, FERMI and ATIC Anomalies}
\author{A. A. El-Zant$^{1}$, S. Khalil$^{1,2}$, and H. Okada$^{1}$\\
$^{1}$ Centre for Theoretical Physics, The British University in
Egypt, El Sherouk City, Postal No, 11837, P.O. Box 43, Egypt \\
$^{2}$ Department of Mathematics, Ain Shams University, Faculty of
Science, Cairo, 11566, Egypt }

\abstract{If dark matter (DM) annihilation accounts for the
tantalizing excess of cosmic ray electron/positrons, as reported
by the PAMELA, ATIC, HESS and FERMI observatories, then the
implied annihilation cross section must be relatively large. This
results, in the context of standard cosmological models, in very
small relic DM abundances that are incompatible with astrophysical
observations.  We explore possible resolutions to this apparent
conflict in terms of non-standard cosmological scenarios;
plausibly allowing for large cross sections, while maintaining
relic abundances in accord with current observations. }


\preprint{}
\begin{document}
\section{Introduction}
The context of our investigation is set by the overwhelming
evidence for a non-baryonic component dominating the matter
content of our Universe, as inferred from a combination of
colluding astrophysical phenomena (the dynamics of galaxies and
clusters; large scale structure in the Universe; cosmic microwave
background fluctuations; big bang nucleosynthesis). Though relying
exclusively on large scale gravitational signatures, the implied
existence of DM alludes to microscopic fundamental physics beyond
the standard model. But although there are several ongoing
attempts at direct detection of expected DM candidates, apart from
the controversial results of DAMA
collaboration~\cite{Bernabei:2008yi} and recent tentative data
points emanating from the CDMSII experiment~\cite{Ahmed:2009zw},
their outcomes have hitherto been invariably negative. It is in
this context that the significant excitement concerning the recent
measurements of excess electron and positron flux, as reported by
the PAMELA, ATIC, HESS and FERMI collaborations, among others,
arose.

 PAMELA's observations, reporting excess flux between $8$
and $80$ {\rm GeV}~\cite{Adriani:2008zr}, with no excess in the
corresponding anti-proton flux, confirm and extend previous
results obtained by HEAT \cite{Barwick:1997ig} and AIMS
\cite{Aguilar:2007yf}. The ATIC~\cite{:2008zzr} experiment data,
on the other hand, show significant excess electron and positron
flux at energies around $300-800$ {\rm GeV}.

Although the PAMELA data is marginally consistent with
calculations of the cosmic ray background employing a 'soft'
background electron flux spectrum~\cite{Delahaye:2008ua}, the ATIC
and FERMI observations more clearly point to the existence of an
additional source of high energy electrons and positrons; and
since such particles cannot travel very far without much energy
loss, their source must be local ($\lsim 1$ kpc from the solar
system). There are plausible astrophysical explanations for this
excess too, e.g. in terms of local pulsars and supernovas
remnants~\cite{Shaviv:2009bu}~\cite{Profumo:2008ms}, but they seem less natural
than when invoked in the low energy region of the spectrum.
The excesses in flux could also result from DM annihilation or
decay, and a vast literature has recently arisen around the
subject (see, e.g., \cite{Cirelli:2008pk}~\cite{Hooper:2009zm} for a
review), partly instigated by a salient feature (perhaps
especially prominent in the ATIC measurements) of the cosmic ray
positron spectrum produced by DM annihilation; the fact that it
drops off at $E_{e^+} = m_{\chi}$, while the flux produced by,
e.g., a single pulsars falls off more gradually. Though this
feature seems significantly more subdued in the FERMI
data~\cite{FERMIDAT}, the relevant region is relatively less well
represented there --- at least in comparison with the HESS data,
where a palpable drop-off is present~\cite{HESS1}, \cite{HESS2}.
Nevertheless, is difficult to conclusively differentiate between
astrophysical and DM descriptions of the observed positron
excesses in any conclusive manner, although future experiments may
be able to achieve this. One advantage of DM models is that,
despite their diversity, their cosmic rays emanate from far
simpler physical objects and are therefore far easier to falsify
than those pertaining to astrophysical sources.

In any case, if the DM scenario is to explain the observed
anomalous flux, one major hurdle has to be surpassed. At present
the major difficulty facing the DM annihilation is the large cross
section apparently required to fit the excess flux --- one that
seems incompatible with straightforward estimates of the relic DM
abundance in conventional cosmological models.

This paper is an attempt at reconciling the aforementioned
observed cosmic ray excess with an interpretation of the results
in terms of annihilating halo dark matter particles. In the next
section we argue that a widely disseminated solutions to this
problem, in terms of boost factors, including those sprouting from
Sommerfeld enhancement and similar effects, seems untenable. From
this stems our motivation for investigating non-standard
cosmological models, allowing for larger DM annihilation cross
sections while being less severely constricted by current
observations.

\section{PAMELA, ATIC, FERMI and HESS Anomalies}
The PAMELA experiment measured an excess of cosmic ray positrons
with no indication of any excess of anti-proton flux. Therefore,
if it is indeed the dark matter that is responsible for the
positron excess, it seems natural to consider a type of DM
particle that annihilates predominantly into $l^+ l^-$ channels.
The positron flux in the galactic halo is given in terms of the
production rate of the
positrons from DM annihilation, which is given by %
\be %
Q(E,{\bf r})= \frac{1}{2}\left(\frac{\rho({\bf
r})}{m_{\chi}}\right)^2 \sum_f \langle \sigma v \rangle_f \left(\frac{dN}{d E}\right)_f,
\label{Qer}%
\ee where $\langle \sigma v \rangle_f \equiv a_f$ refers to the
averaged annihilation cross section into the final state $f$ and
$(dN/dE)_f$ is the fragmentation function, representing the number
of positrons with energy $E$, produced from the final state $f$.
$\rho ({\bf r})$ is a DM halo mass profile. Though there are
several types of proposed halo dark matter density profiles, we
adopt here the standard NFW profile \cite{navaro}.
\FIGURE{
\epsfig{file=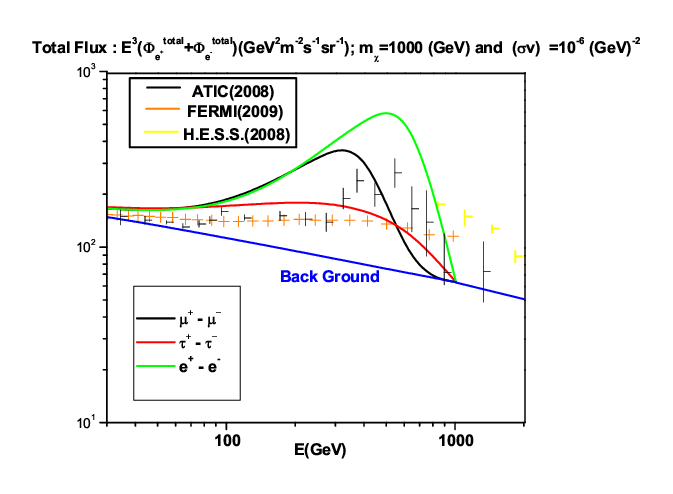, width=8.5cm,height=4.9cm,angle=0}
\caption{The total absolute flux in units of ${\rm
GeV^2m^{-2}s^{-1}sr^{-1}}$, generated by the DM annihilation into
$e^+e^-$(green line), $\mu^+ \mu^-$(black line) and $\tau^+
\tau^-$(red line), as function of positron energy for $m_{\chi}
=1$ ${\rm TeV}$, for thermal averaging cross
section 
$10^{-6}~ {\rm GeV}^{-2}$.}
\label{flux}
}

Fig.~\ref{flux} shows the total absolute flux,
$\Phi_{e^+}^{total}+\Phi_{e^-}^{total}$, generated by the DM
annihilation into $e^+ e^-$, $\mu^+ \mu^-$ and $\tau^+ \tau^-$ as
function of positron energy for $m_{\chi} =1$ {\rm TeV} and
$10^{-6}~ {\rm GeV}^{-2}$. In our analysis, the MED diffusion
model of Delahaye et. al.~\cite{donato} and NFW\cite{navaro}
Galactic halo with scale-length 20 kpc are assumed. Also,
$\Phi_{e^+}^{total}$ and $\Phi_{e^-}^{total}$ are defined as \be
\Phi_{e^+}^{total}\sim\Phi_{e^+}^{DM}+\Phi_{e^+}^{sec}\quad,\quad
\Phi_{e^-}^{total}\sim\Phi_{e^-}^{prim}+\Phi_{e^-}^{sec}.\ee

The flux of positrons$\Phi_{e^+}^{DM}$ is given from the number
density of positron through several steps(for instance, see ref.
\cite{Hisano:2005ec}).

The fragmentation function for direct process is almost monotonic;
$\L(\frac{dN}{dE}\R)_{ee}\sim \delta(E-m_{\chi})$. For
 $\L(\frac{dN}{dE}\R)_{\mu\mu}$ and $\L(\frac{dN}{dE}\R)_{\tau\tau}$,
we referred to the ref. \cite{Pallis:2009ed}.

Calculated in terms of fitting functions matching the fluxes
deduced via standard simulations of cosmic ray production and
propagation~\cite{baltz}~\cite{dela.}, the astrophysical
background fluxes of positrons; $\Phi_{e^+}^{sec}$,
$\Phi_{e^-}^{prim}$ and $\Phi_{e^-}^{sec}$, are given by the
followings:\bea \Phi^{prim.}_{e^-}(\epsilon) &=&\frac{0.16
\epsilon^{-1.1}}{1+11 \epsilon^{0.9}+3.2
\epsilon^{2.15}}(cm^{-2}s^{-1}sr^{-1}), \nn\\
\Phi^{sec.}_{e^-}(\epsilon) &=& \frac{0.70 \epsilon^{-0.7}}{1+110
\epsilon^{1.5}+600 \epsilon^{2.9}+580
\epsilon^{4.2}}(cm^{-2}s^{-1}sr^{-1}),
\nn\\
\Phi^{sec.}_{e^+}(\epsilon) &=& \frac{4.5 \epsilon^{0.7}}{1+650
\epsilon^{2.3}+1500\epsilon^{4.2}}(cm^{-2}s^{-1}sr^{-1}), \eea
where $\epsilon\equiv E/(1 {\rm GeV})$.

We use a DM particle of mass 1 {\rm TeV}, consistent with the
maximum allowed by FERMI $\gamma$-ray observations of dark matter
dominated virilized gravitational systems~\cite{FERMIclus},
\cite{FERMIdwardf}.
As is immediately apparent, $\Phi_{e^+}^{sec}$ and
$\Phi_{e^-}^{sec}$ are quite small at all energies, and can be
neglected with respect to $\Phi_{e^-}^{prim}$; the same goes for
the computed positron flux for $\langle \sigma v \rangle \sim
10^{-9} ~{\rm GeV}^{-2}$, which remains far below the background.
Therefore, in such a situation, and in terms of the plotted
quantities, any excess flux would have to be explained in terms of
modifications to the astrophysical background (including the
possible effects of local pulsars, supernovae remnants etc.).
Alternatively, a larger cross section can lift the computed DM flux
above the background, so as to explain any apparent excess. It
turns out that a $\langle \sigma v \rangle \sim 10^{-6} ~{\rm
GeV}^{-2}$ is compatible with PAMELA observations (without
invoking any changes in the background model). In addition, for positron
energy $E
> 200$ {\rm GeV}, the positron flux $\Phi_{e^+}^{DM}$ exceeds the electron
background, which fares well in explaining the ATIC, HESS and
FERMI excess at $300-800$ {\rm GeV}.

\FIGURE{
\epsfig{file=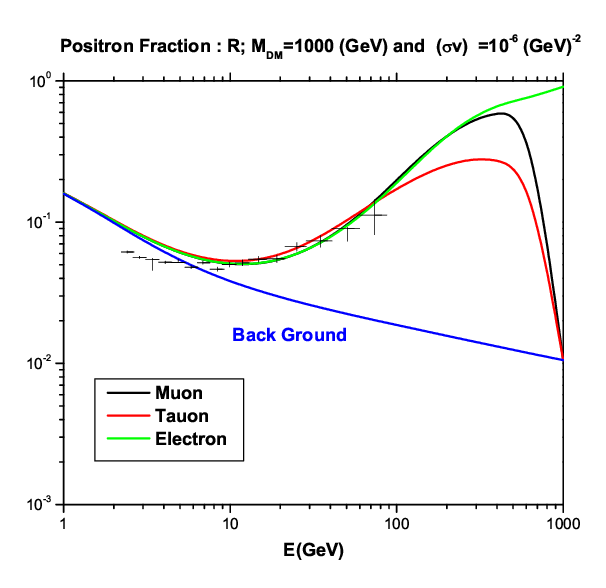, width=8.5cm,height=4.9cm,angle=0}
\caption{The positron fraction for DM annihilation into $e^-
e^+$(green line), $\mu^+ \mu^-$(black line) and $\tau^+
\tau^-$(red line) for dark matter mass $m_{\chi} =1$ {\rm TeV} and
thermal averaging cross section $10^{-6}~ {\rm GeV}^{-2}$.}
\label{fraction}
}

In Fig.~\ref{fraction}, we plot the positron fraction
\be%
R = \frac{\Phi_{e^+}^{DM}(E) +
\Phi_{e^+}^{sec}(E)}
{\Phi_{e^+}^{DM}(E) + \Phi_{e^+}^{sec}(E)+
\Phi_{e^-}^{prim}(E) + \Phi_{e^-}^{sec}(E)},%
\ee the quantity that PAMELA actually measured, for DM
annihilation into $e^+e^-$, $\mu^+ \mu^-$ and $\tau^+ \tau^-$,
along with the relevant measurements. It is clear, as expected,
that DM annihilation with $\langle \sigma v \rangle \sim 10^{-6} ~
{\rm GeV}^{-2}$ can easily account for the PAMELA measurements
while employing a standard astrophysical background. These
measurements explore the existence of positron excess for energies
confined between $8$ and $80$ {\rm GeV}. In the next PAMELA
experiment, the search for excess flux will be extended up to
energies $\sim 270$ {\rm GeV}. We note here that, at such
energies, the positron flux due to DM annihilation is of order the
electron primary flux and so the expected positron fraction would
tend toward
\be %
R \simeq
\frac{\Phi_{e^+}^{DM}}{(\Phi_{e^-}^{prim}+\Phi_{e^-}^{DM}) +
\Phi_{e^+}^{DM}} \simeq {\cal O}(0.3). %
\ee
Such a result would represent a significant signature confirming
that DM annihilation is the source of this observed positron
excess, making the case for an increased DM flux magnitude
(compared to what can be inferred from calculations using standard
smooth halo density profiles and cross sections compatible with thermal
relic abundance calculations), particularly  more pressing.

Already, myriads of remedies, invoking various 'boost factors',
have been proposed. Including, for example, in terms of increased
DM density inside the galaxy's the subhalo population. Yet very
little such substructure is expected in the solar neighborhood;
efficient stripping dissolves subhalos in these inner regions,
ensuring that most mass in subhalos is concentrated at far larger
galactocentric radii. And positrons with energies of order $100$
${\rm GeV}$ or above are expected to emanate from DM annihilations
within a few hundred pc of the solar system. This is quite a
generic result, independent of the details of diffusion model
adopted for cosmic ray propagation, having its origins in such
fairly tractable processes, such as the expected positron energy
loss due to inverse Compton scattering off cosmic microwave
background~\cite{Slatyer:2009yq} and starlight. It is therefore quite unlikely that the
required boost in DM annihilation flux can be obtained by invoking
the galactic subhalo population --- a result confirmed by detailed
modelling of its effect~\cite{lavalle:0902}~\cite{athanass}.

While the existence of other density-borne boosts to the flux
cannot, in principle, be completely ruled out (e.g., in addition to the subhalo
population, DM caustics, tidal streams, mini-black holes hosting
DM spikes etc. have been proposed), a large local enhancement is
unlikely to equally boost the positron flux at all energies, and so
leave the fits to the flux invariant --- again because of the energy dependent
manner in which positrons lose energy as they move through the
galaxy; indeed, the case for streams and caustics seems to  have been falsified, even
in principle, i.e. in terms of source function in~\cite{white:10}.
The process of fitting the data therefore requires
significant fine tuning of masses and positions of objects (as is also
the case when invoking the standard subhalo population~\cite{hooper:zurek}).
Many of these proposals are, in addition, in significant conflict with, or highly
constrained by, $\gamma$ ray observations~\cite{bring:0902} \cite{Bertone:2008xr}.

  Another class of proposals involve velocity dependent cross sections
originating from Sommerfeld enhancement effects
(e.g., \cite{silk}~\cite{arkani}~\cite{Decourchelle}~\cite{Hisano:2004ds}).
These can dramatically increase annihilation rates at velocities
characteristic of galactic halos without affecting the relic
abundance. This boost however comes at the price of fine tuning
the particle masses~\cite{Decourchelle}; it requires new light
scalar or gauge boson, $\phi$, to mediate the annihilation of DM
into leptons. The mass of this particle is constrained as:
\be
m_{\phi} \lsim \alpha m_{\chi} \lsim {\rm few~GeV},\label{mass-bound}
\ee
where $\alpha$ refers to the coupling of the $\phi$ interaction squared over
$4\pi$.

   Furthermore, proposals for further enhancing the cross
section for annihilation by considering the still lower relative
particle velocities inside distant subhalos~\cite{silk}, or in the
central regions of halo cusps that are shallower than
isothermal~\cite{zentner}, are untenable, because, again, high
energy positrons originating in these regions cannot safely
traverse the space separating them from the solar neighborhood.

  In addition,Cosmic Microwave background constraints seem
to rule out dramatic enhancements due to low dark velocities
matter velocities in subhalos, since they suggest the low-velocity
annihilation enhancement must have saturated at the last
scattering surface (when $v_{\rm DM}/c \sim 10^{-8}$). WMAP
results do not, on the other hand, rule out enhanced cross
sections consistent with dark matter annihilation from the main
halo that explain the cosmic ray excesses. They therefore do not
falsify the models described below~\cite{slatyer:09} (though
PLANCK results may).  Finally, FERMI $\gamma$-ray constraints also
seem to rule out enhanced low particle velocity enhanced boosts in
subhalos~\cite{FERMIclus}\cite{FERMIdwardf}.


\section{Relic Abundance and Non-conventional Cosmology}

During the radiation epoch in the early Universe, if the DM is
assumed to be in thermal equilibrium, at the freeze out
temperature $T_F$, the standard calculation for the thermal
average of the annihilation cross section, for $v \ll c$, yields: %
\be
\langle \sigma v \rangle_{F} = a + b~ v_{F}^2,%
\ee %
where $v_{F}$ is the velocity of the DM at the freeze out
temperature, which is of order $0.1~c$. Note that the above
expansion of $\langle\sigma v \rangle$ is consistent only far from
s-poles and threshold. In our galactic halo the velocity of the DM
particles is much smaller (of order $10^{-3} \times c$).  Its
distribution is also no longer Maxwellian and its dispersion may
vary significantly with radius. Nevertheless, due to the
suppression of the kinetic energy of the DM  respect to its rest
mass, the average cross section in the galactic halo has
negligible dependence on the velocity distribution function (we
have explicitly checked this point by considering the isotropic
distribution function \cite{midrow} associated with the NFW
density profile). In this context, $\langle
\sigma v \rangle_{halo}$ can be written as %
\be %
\langle \sigma v \rangle_{halo} \simeq a .%
\ee %
This implies that the value of $\langle \sigma v \rangle$,
involved in the computation of the positron flux in the galactic
halo, is the $s$-wave annihilation, which contributes to the DM
relic abundance $\Omega h^2$.

   Given the difficulties outlined above, it seems pertinent to ask whether
the constraints imposed on the cross section by standard relic
abundance calculations are as universally prohibitive as they seem; and to
examine mechanisms whereby they could be circumvented.

  The usual assumption is that the DM particles decoupled from
the standard model particles when the former became
non-relativistic. In that case the DM relic density is given by %
\be%
\Omega_{\chi} h^2 = \frac{8.76 \times 10^{-11} {\rm
GeV}^{-2}}{g_*^{1/2}(T_F) \left(a/x_F + 3 b/x_F^2\right)},%
\ee
where $T_F$ is the freeze out temperature, $x_F = m_{\chi}/T_F
\simeq 20$, and $g_*$ the number of relativistic degrees of
freedom. It is clear that, for $a \sim 10^{-6}$ $GeV^{-2}$, one
gets $\Omega_{\chi} h^2 \sim 10^{-4}$, which is inconsistent with
the WMAP results \cite{Hinshaw:2008kr}, as well as a host of
observations concerned with the dynamics and large scale
distribution of galaxies and clusters.

It is nevertheless important to note that relaxing this
assumption can very well give rise to different predictions for
the relic density without affecting any of the relevant
observations. We will focus here on two such non-standard
cosmological scenarios as examples illustrating this point, thus
resolving the tension between the relic abundance constraints and
the PAMELA and ATIC results.
Our first example considers a cosmological scenario where the
reheating temperature is associated to the decay of a standard
model singlet field $\psi$, in which case the DM relic density
considerably increases with respect to the standard
radiation-dominated case, by virtue of the effect of direct
non-thermal production by the $\psi$ field \cite{Khalil:2002mu}.
The second example is based on the non-conventional brane
cosmology \cite{Abou El Dahab:2006wb}, which also allows for
several orders of magnitude increase in the cross section.
The principal point in this first example concerns the numerical
relation between the reheating temperature $T_{RH}$ and the the DM
freezing temperature $T_F$. The former being defined as the
temperature at which the oscillating field energy ceases to
dominate the cosmological evolution, heralding the start of the
radiation dominated epoch. In the standard scenario, where $T_{RH}
\sim 10^9$ ${\rm GeV}$, the reheating epoch has no relevance in
the final output of the DM relic density. However, for low value
of $T_{RH}$, so that $T_{RH} < T_F$, it can have important
implications on the predictions of the relic abundance of DM as
discussed in Ref.\cite{Khalil:2002mu}.

Thus, in this context,  it is possible to envision the existence
of an epoch in the history of the Universe, preceding the
radiation-dominated era, when the energy density was dominated by
coherent oscillating fields; the so called reheating era. The
decay width of this scalar field $\psi$ can be
parameterized via %
\be%
\Gamma_{\psi} = \frac{1}{2\pi} \frac{m_{\psi}^3}{M_*^2}, %
\ee
where $M_*$ defines a high scale, acting as an effective suppression
scale.

The coupled Boltzmann equation, for the DM, scalar field $\psi$
and radiation, is then solved. The detailed equations used in this
computation can be found in Ref.~\cite{Khalil:2002mu}. The
resulting relic density can then be estimated to be \cite{Moroi:1999zb} \cite{Mizuta:1992qp}%
\begin{eqnarray}%
\Omega_{\chi} h^2 \sim \left(\frac{M_*}{1.5 \times 10^{20} {\rm
GeV}}\right) \left(\frac{1~{\rm GeV^{-2}}}{\langle \sigma v
\rangle}\right)\left(\frac{100~{\rm GeV}}{m_{\psi}}\right)^{3/2}
\left(\frac{10.75}{g_*}\right)^{1/4} \left(\frac{m_{\chi}}{100{\rm GeV}}\right). %
\end{eqnarray}%
From this equation, it can be seen that it possible to obtain
$\Omega_{\chi} h^2 \sim {\cal O}(0.1)$ with an annihilation cross
section of order $10^{-6}~ {\rm GeV}^{-2}$, as required by PAMELA
and ATIC data. For example, for $m_{\psi} = 1$ ${\rm TeV}$ and
$M_* = 10^{14}$ ${\rm GeV}$ we obtain $\Omega_\chi h^2 \simeq
0.1$.

\FIGURE{
\epsfig{file=RM5.eps, width=8.5cm,height=4.9cm,angle=0}
\caption{The enhancement/suppression factor
$R=(\Omega_{\chi}h^2)_b/(\Omega_{\chi} h^2)_s$ as a function of
the five dimensional scale $M_5$ (${\rm GeV}$) for $m_{\chi} =100$
(solid curve), $200$ (dashed curve) and $500$ ${\rm GeV}$ (dotted
curve).}
\label{RM5}
}
We now turn to the second possible scenario invoking non-standard
cosmology as a mechanism for boosting the DM annihilation cross
section. As mentioned above, it is based on the brain world
cosmology, which is embedded in five dimensional warped space
time. In this case, the derived Friedman equation is given by %
\be%
H^2 = \frac{8 \pi G_{(4)}}{3} \rho \left( 1 + \frac{\rho}{2
\sigma}\right) - \frac{k}{a^2}+ \frac{{\cal C}}{a^4}, %
\ee
where $H=\dot{a}/a$ is the Hubble parameter and $a(t)$ is the
scale factor, $\rho$ is the energy density of ordinary matter on
the brane, while $\sigma$ is the brane tension; $G_{(4)}$ refers to
the $4D$ Newton coupling constant, $k$ stands for the
curvature of our three spatial dimensional and ${\cal C}$ is a
constant of integration known as dark-radiation.
This equation implies that $H \propto \rho$ rather than
$\sqrt{\rho}$ as in the standard cosmology. Thus, the evolution of
the scale factor will be different from the usual one.
This modification affects any relic abundance of DM that may be reminiscent of
the radiation dominated phase of the early Universe
~\cite{Abou El Dahab:2006wb}~\cite{Randall:1999vf}~\cite{Okada:2004nc}.
For, in this model, the Universe undergoes a nonstandard brane
cosmology at early times till it reaches a temperature, known as
transition temperature $T_t$, when it sustains the
standard cosmology. This transition temperature is defined as %
\be %
\rho ({T_t})= 2\sigma ~~ \Rightarrow ~~ T_t = 0.51 \times 10^{-9}
M_5^\frac{3} {2}~\rm{GeV}, %
\ee %
and the transition should take place above the nucleosynthesis era
({\it i.e.}, $T_t > 1$ \rm{MeV}). Here $M_5$ is the five
dimensional Plank mass. If the freeze out temperature of the DM
$(T_F)$ is higher than the transition temperature, {\it i.e.},
$T_F \geq T_t$ and $M_5 \leq 10^5$ \rm{GeV}, one finds that the
ratio between the relic density in brane and standard cosmology is
given by \cite{Abou El Dahab:2006wb}
\be %
 R = (\Omega_{\chi}h ^2)_b/(\Omega _\chi h^2)_s \simeq {\cal
O}(10^2 - 10^{3}).%
\ee %
In Fig. \ref{RM5}, we present the prediction for the factor $R$ as
a function of the scale of the five dimensions, $M_5$, for
different values of $m_{\chi}$, namely we consider $m_{\chi} =
100, 200$ and $500$ ${\rm GeV}$. As can be seen from this figure,
for $M_5 < 10^6$ the brane cosmology effect is quite large and the
factor $R$ becomes much larger than one. In this case the
resulting relic density $(\Omega_{\chi} h^2)_b$ may exceed the
WMAP results $ \Omega_{\chi}h^2 \simeq 0.1$. Moreover for $M_5
\gsim 5 \times 10^{6}$, the ratio $R$ becomes less than one and a
small suppression for $(\Omega_{\chi} h^2)_s$ can be obtained.

 An analogous analysis, in the case of standard cosmology, would give
$(\Omega h^2)_s\simeq 10^{-3}$, under the fixed parameters
$m_{\chi}=1$ $TeV$, $\langle\sigma v\rangle=10^{-6}$. Hence, 
for this calculation to be consistent with  $(\Omega h^2)_b\simeq 0.1$,
$R$ must reach at around $10^2$. Hence $M_5$, in our case, has to
be more than $10^6$ from the Fig. \ref{RM5}.

This brane enhancement or suppression for the dark matter relic
density could be favored or disfavored based on the value of the
relic abundance in the standard scenario. If
$(\Omega_{\chi}h^2)_s$ is already larger than the observational
limit, as in the case of bino-like particle, then a suppression
effect would be favored and hence $M_5$ is constrained to be
larger than $5 \times 10^{6}$ ${\rm GeV}$. However, for wino- or
Higgsino-like particle where the standard computation usually
leads to very small relic density, the enhancement effect will be
favored and the constraint on $M_5$ can be relaxed a bit
\cite{Nihei:2005qx}. In general, it is remarkable that in this
scenario the dark matter relic density imposes a stringent
constraint on the fundamental scale $M_5$. This brane enhancement
implies that stable particles with annihilation cross section
$\sim 10^{-6}~ {\rm GeV}^{-2}$, which seems essential for
accommodating the PAMELA and ATIC results, remain viable DM
candidates.


\section{Extra Constraints on the Annihilation Cross Section and the Viability of our Model}

For our model to be viable, it has to be consistent with other
constraints. Some of these were briefly mentioned earlier in our
discussion; we summarize the situation as regards to these here.

DM annihilation at redshift $z \sim 1000$ has an effect on the
degrees of ionization of the primordial plasma, by virtue of the
energy it pumps into it through the high energy particles it
produces. By affecting  its degree of ionization, it broadens the
last scattering surface, which in turn has an effect on
temperature correlations and polarization of the cosmic microwave
background. Slatyer {\em et. al.}~\cite{slatyer:09} have presented
a detailed study of the phenomenon. Their results tend to disfavor
the large boosts implied by Sommerfeld enhancements in dark matter
substructure, but not as much the standard enhancement at normal
local dark halo velocities.
 WMAP results therefore do not rule out the cross section boosts required for fitting the
PAMELA and ATIC results in the context of non-standard cosmological scenarios,
and are even less severe in constraining
the FERMI data.
Though further constraints from PLANCK would provide further stringent tests
that may in principle falsify the scenarios presented here.

  Since dark matter annihilation produces secondary gamma rays,
through internal bremsstrahlung and inverse Compton scattering,
there are also severe constraints on DM annihilation cross
sections from the observed gamma ray fluxes; in particular from
recent observations by FERMI .The most stringent appear to be
those concerning the gamma rays emanating dark matter dominated
systems such as dwarf galaxies and galaxy clusters.Recently
published results from eleven months of FERMI
observations~\cite{FERMIclus}, \cite{FERMIdwardf}, seem to
disfavor cross sections consistent with the observed cosmic ray
excesses if the particles producing these fluxes exceed the ${\rm
1 TeV}$ limit, particularly if the annihilation proceeds through
the muon channel. Thus all the models presented here are (albeit
in one case marginally) consistent with these constraints.

Finally, the current experimental limits from CDMS~\cite{cdms} and
XENON10~\cite{xenon10} imply that the DM-nucleus scattering cross
section is less than $10^{-7} ~ {\rm GeV}^{-2}$. By crossing
symmetry in Quantum Field Theory (QFT), the annihilation cross
section and the scattering cross section are related; which may
lead one to suspect the presence of a contradiction between the
direct detection results and the attempt to explain the PAMELA and
ATIC data via enhanced annihilation cross sections. But this
apparent conflict can be resolved by assuming that the DM particle
has suppressed coupling with quarks, while retaining reasonably
large coupling with leptons; {\it i.e.}, that the DM has a
leptonic nature. As mentioned above, this type of DM is also
favored for explaining the absence of flux excess in the observed
anti-proton flux in PAMELA, ATIC and FERMI data. It is clear that
in this case one can naturally have the annihilation cross section
into $l^+ l^-$ of order $10^{-6} ~ {\rm GeV}^{-2}$, whereas the
scattering cross section of DM and nucleus $\sigma_{\rm DM-N}$ is
quite suppressed. 

It is also important to note that an $s$-wave dominated
annihilation cross section of order $10^{-6}~{\rm GeV}^{-2}$ bears
interesting corollaries concerning the nature of prospective DM
candidates, and by implication on particle physics theory. It is
well known, for example, that if the DM is composed of Majorana
particles, then its annihilation cross section into leptons is
proportional to $m_l^2/m_{\chi}^4$; which, for $m_{\chi}\simeq
100$ {\rm GeV}, is typically less than $10^{-8}~{\rm GeV}^{-2}$.
However, for Dirac or scalar boson type DM the annihilation cross
section is no longer proportional to the lepton mass squared and
it may thus be enhanced significantly. In this context, the famous
DM candidate of lightest neutralino in minimal supersymmetric
standard model, which is a Majorana particle, is not favored for
explaining the PAMELA/ATIC results. On the other hand, the
right-handed sneutrino, which is considered an interesting scalar
candidate for DM in supersymmetric theories with right-handed
neutrino~\cite{Arina}, can account for the large annihilation
cross sections required for explaining the PAMELA/ATIC
measurements.


\section{Conclusions}

In this paper we have proposed solutions reconciling the
interpretation of the observed cosmic ray excesses with dark
matter annihilation models by invoking non-standard cosmologies.
We have argued that these models have several advantages over
those invoking density or velocity borne boost factors, especially
those invoking large enhancements due to dark matter overdensities
and halo substructures. At present, our models are consistent with
current observations. Future data, especially those from the
PLANCK concerning the enhanced effect of DM annihilation on the
Cosmic Microwave Background, will determine whether our models
remain viable.

\section*{Acknowledgments}
This work was partially supported by the ICTP Project ID 30,
Science and Technology Development Fund (STDF) Project ID 437, and
the Egyptian Academy for Scientific Research and Technology.

\end{document}